\documentclass[11pt]{JHEP3}

\usepackage{epsfig}
\usepackage{citesort}

\title{\boldmath The Forward Backward asymmetries of $B \to X_s \tau^+
\tau^-$ in the MSSM}

\author{A. S. Cornell \\
  Korea Institute for Advanced Study, 207-43
Cheongryangri 2-dong, Dongdaemun-gu, Seoul 130-722,\\
 E-mail: \email{alanc@kias.re.kr}}
\author{Naveen Gaur \\
  Department of Physics \& Astrophysics, University of Delhi, \\
  Delhi - 110 007, India \\
  E-mail: \email{naveen@physics.du.ac.in}}

\abstract{
The relatively clean theoretical probes of the Standard Model
(SM), and the various theories beyond the SM, provided by
radiative, semi-leptonic and (purely) leptonic decays of B-mesons
have become increasingly important. Due to the large number of
possible distributions in the semi-leptonic decays based on the
quark level transition $b \to s \ell^+ \ell^-$ (not just the
branching ratio), these transitions have become very useful. A
study of the Forward-Backward asymmetries for the inclusive decay
($B \to X_s \ell^+ \ell^-$) is carried out in this paper.  This
study shall be performed in the SM and a minimal supersymmetric
extension of the SM, namely the mSUGRA model.
}

\preprint{}


\begin{document}


\section{\label{section:1}Introduction}

The recent flux of papers studying the various decay processes of
B-mesons stands testimony to the great importance of this field in
testing not only the parameters of the Standard Model (SM) but in
constraining the parameters of its many possible extensions.  In
this vain processes of particular interest are the so called
``{\em rare}" decays, these being transitions through loop levels
only \cite{Buchalla:1995vs}. Of particular interest are the
Flavour Changing Neutral Currents (FCNC) $b \to s(d)$. These rare
decays are extremely sensitive to the masses and couplings of the
virtual particles involved in this transition, and therefore
provide the greatest hope of determining many of the SM's (and its
possible extensions) underlying structures.  We therefore require
a decay process which is both theoretically clean and
experimentally realizable in the near future, such as $B \to X_s
\gamma$\footnote{this decay mode has already been observed in
various B-factories and has proved to be extremely useful in
placing stringent bounds on many new physics models} and $B \to
X_s \ell^+ \ell^-$.  Of these two processes this paper shall focus
on the inclusive decay mode, $B \to X_s \ell^+ \ell^-$, for the
reason that this decay mode is a far richer alternative to the
radiative mode as here many distributions involving the final
state lepton pair can be measured. In the inclusive decay channel
($B \to X_s \ell^+ \ell^-$) observables such as Forward-Backward
(FB) asymmetries and lepton polarization asymmetries, where one or
both the leptons are polarized, have been extensively studied both
within the SM and its various extensions
\cite{Gaur:2003dy,RaiChoudhury:1999qb,Kruger:1996cv}.

\par Recently, it has been pointed out by Bensalem et al.
\cite{Bensalem:2002ni} that one can construct the polarized FB
asymmetries also (FB asymmetries when one or both the leptons are
polarized). These new observables could provide a sufficiently
large number of observables for a very strict testing of the SM
and any new physics. These polarized FB asymmetries were studied
in an exclusive decay mode ($B \to K \ell^+ \ell^-$) both within
the SM and the Minimal Supersymmetric SM (MSSM)
\cite{Choudhury:2003xg}, motivated by the vanishing unpolarized FB
asymmetry for $B \to K \ell^+ \ell^-$ within the SM
\cite{Bobeth:2001sq,Iltan:1998a}. Based on the construction of
doubly polarized observables by Bensalem et al. such observables
were studied in the case of $B \to K^* \ell^+ \ell^-$
\cite{Choudhury:2003mi} and in the inclusive mode \cite{Gaur:2003dy}.
In this work we will focus on the polarized FB asymmetries.

\par Lately there has been an increased interest in how the predictions
of various observables will change in the two Higgs doublet model
(2HDM) and Supersymmetric (SUSY) extensions of the SM.  The
primary reason for this is due to the increase in the number of
particles in these theories, such as the Neutral Higgs Bosons
(NHBs), and hence scalar and pseudo-scalar exchange operators are
now included.  Such operators give rise to many interesting new
possibilities, like orders of enhancement in the branching ratio
for purely leptonic decays ($B_{s,d} \to \ell^+ \ell^-$)
\cite{Choudhury:1999ze,Skiba:1993mg} and non-vanishing values for
unpolarized FB asymmetries in $B \to K(\pi) \ell^+ \ell^-$
\cite{Choudhury:2003xg,Bobeth:2001sq,Iltan:1998a}. As the coupling
of the NHBs to the leptons is proportional to the lepton mass
\cite{Choudhury:1999ze,Xiong:2001up,RaiChoudhury:1999qb,Skiba:1993mg},
the effects of these new particles is greater when the final state
leptons are either $\mu$ or $\tau$. Therefore in this paper we
will also consider the effect of the new scalar and pseudo-scalar
operators which arise from a MSSM. The particular model to be
employed shall be explained in more detail in section
\ref{section:4}.

\par As such, this paper shall therefore be organized as follows:  In
section \ref{section:2} we shall review the effective Hamiltonian
and the matrix element for the process concerned, as derived from the
quark level transition $b \to s \ell^+ \ell^-$.  Section
\ref{section:3} will then focus on the definitions and analytic
results of the FB asymmetries.  Finally, section \ref{section:4} will
contain our discussion and numerical analysis of the results.


\section{\label{section:2} Effective Hamiltonian}
The effective Hamiltonian for the quark level transition $b \to s
\ell^+ \ell^-$, as first studied by Grinstein et al. (and later to
NLO by Buras and M\"{u}nz) \cite{Grinstein:1989me}\footnote{the
complete NNLL calculation of $B \to X_s \ell^+ \ell^-$ has
recently been given in reference \cite{Ghinculov:2002ge}, however
for our analysis we will use only the Leading Log results given in
reference \cite{Grinstein:1989me}.}, and represents the quark
level process for the inclusive decay $B \to X_s \ell^+ \ell^-$,
under consideration (where we shall use descriptions as penned in
references
\cite{Choudhury:1999ze,Xiong:2001up,RaiChoudhury:1999qb});
\begin{equation}
{\cal H}_{eff} = \frac{4 G_F}{\sqrt{2}} V_{tb}V^*_{ts} \left(
\sum_{i=1}^{10} C_i O_i + \sum_{i=1}^{10} C_{Q_i} Q_i \right)
\label{sec2:eq:1}
\end{equation}
where the $O_i$ operators are the current-current (i=1, 2), penguin
(i=3,\dots,6), magnetic penguin (i=7,8) and semi-leptonic (i =
9,10) operators with the $C_i$ corresponding to the standard
Wilson coefficients.  The additional operators $Q_i$ (i =
1,\dots,10) and their Wilson coefficients ($C_{Q_i}$) are related
to the additional particles of the MSSM under consideration.

\par Using the above effective Hamiltonian, and neglecting the s-quark
mass, the matrix element for the quark level transition $b \to s
\ell^+ \ell^-$ is;
\begin{eqnarray}
{\cal M} &=& \frac{\alpha G_F}{\sqrt{2} \pi} V_{tb}V^*_{ts}
\Bigg\{ - 2 C_7^{eff} \frac{m_b}{q^2} (\bar{s} i \sigma_{\mu \nu}
q^\nu P_R b) ( \bar{\ell} \gamma^\mu \ell ) + C_9^{eff} (\bar{s}
\gamma_\mu P_L b) (\bar{\ell} \gamma^\mu \ell) \nonumber \\
&& \hspace{2cm} + C_{10} (\bar{s} \gamma_\mu P_L b) (\bar{\ell}
\gamma^\mu \gamma_5 \ell) + C_{Q_1} (\bar{s} P_R b) (\bar{\ell}
\ell) + C_{Q_2} (\bar{s} P_R b) (\bar{\ell} \gamma_5 \ell) \Bigg\}
. \label{sec2:eq:2}
\end{eqnarray}
In this expression $q$ represents the momentum transferred to the
lepton pair, given as $q = p_+ + p_-$, where $p_+$ and $p_-$ are
the momentas of the $\ell^+$ and $\ell^-$ particles respectively.
Note also that the CKM factors have been denoted $V_{tb}V^*_{ts}$
above, and that $P_{L,R} = (1 \mp \gamma_5)/2$.

\par In this paper we will also include the long range effects
which arise from the four quark operators $\langle \ell^+ \ell^-
s| O_i |b \rangle$ for $i = 1,\dots,6$.  This shall be done by
absorbing them into the description of the Wilson coefficients; a
prescription which has been used in many earlier works, see
references \cite{Kruger:1996cv,Long-Distance}.  These long
distance effects typically arise from $c\bar{c}$ contributions,
and are taken into account by adding a term to the $C_9^{eff}$
coefficient;
\begin{equation}
C_9^{res} \propto \kappa \sum_{V = \psi} \frac{\hat{m}_V Br(V \to
\ell^- \ell^+) \hat{\Gamma}^V_{total}}{\hat{s} - \hat{m}_V^2 + i
\hat{m}_V \hat{\Gamma}^V_{total}} \label{sec2:eq:3}
\end{equation}
where we have used the same symbols and notation as in Kr\"{u}ger
and Sehgal \cite{Kruger:1996cv}.  In the above equation the
phenomenological factor $\kappa$ is introduced to predict the correct
branching ratio of $Br(B \to J/\Psi X_s \to X_s \ell^+ \ell^-)$. For
our numerical analysis we will choose $\kappa$ to have a value of 2.3.

\par Using the expression of matrix element in equation
(\ref{sec2:eq:2}) we obtain the expression for the differential decay
rate as;
\begin{equation}
\frac{d \Gamma}{d \hat{s}} = \frac{G_F m_b^5}{192 \pi^3}
\frac{\alpha^2}{4 \pi^2} |V_{tb} V_{ts}^*|^2 (1 - \hat{s})^2
\sqrt{1 - \frac{4 \hat{m}_{\ell}^2}{\hat{s}}} \bigtriangleup
\label{sec2:eq:8}
\end{equation}
with
\begin{eqnarray}
\bigtriangleup &=& 4 \frac{(2 + \hat{s})}{\hat{s}} \left(1 +
\frac{2 \hat{m}_{\ell}^2}{\hat{s}}\right) |C_7^{eff}|^2 + (1 + 2
\hat{s}) \left(1 + \frac{2 \hat{m}_{\ell}^2}{\hat{s}}\right)
|C_9^{eff}|^2 \nonumber \\
&& + (1 - 8 \hat{m}_{\ell}^2 + 2 \hat{s} + \frac{2
\hat{m}_{\ell}^2}{\hat{s}}) |C_{10}|^2 + {3 \over 2} (-4
\hat{m}_{\ell}^2 + \hat{s}) |C_{Q_1}|^2 + {3 \over 2} \hat{s}
|C_{Q_2}|^2 \nonumber \\
&& + 12 (1 + \frac{2 \hat{m}_{\ell}^2}{\hat{s}}) Re(C_9^{eff *}
C_7^{eff}) + 6 \hat{m}_{\ell} Re(C_{10}^* C_{Q_2}) .
\label{sec2:eq:9}
\end{eqnarray}
Using this expression of the invariant mass spectrum (including
the scalar exchange effects) we will now analyze the various FB
asymmetries in the next section.


\section{\label{section:3} Polarized FB asymmetries}
As was mentioned in the introduction, the additional intricacies
of the $b \to s \ell^+ \ell^-$ quark level transition requires
many more experimental observables in processes such as $B \to X_s
\ell^+ \ell^-$ in order to constrain all the parameters of a
particular MSSM.  With this in mind we shall now define the
differential FB asymmetry as \cite{Long-Distance};
\begin{equation}
\overline{{\mathcal A}}(\hat{s}) = \int_0^1 \frac{d^2
\Gamma}{d\hat{s} dz} dz   - \int_{-1}^0 \frac{d^2 \Gamma}{d\hat{s}
dz} dz .
\label{sec3:eq:1}
\end{equation}
If we do not sum over the spins of the outgoing leptons
then the FB asymmetry will, in general, be a function of the spins
of the final state leptons; and can be defined as
\begin{equation}
\overline{{\mathcal A}}(s^-,s^+,\hat{s}) = \int_0^1 \frac{d^2
\Gamma (s^-,s^+)}{d\hat{s} dz} dz   - \int_{-1}^0 \frac{d^2 \Gamma
(s^-,s^+)}{d\hat{s} dz} dz .
\label{sec3:eq:2}
\end{equation}
As our eventual aim is to derive expressions to be searched for in
experiments it is convenient to use the normalized FB asymmetry.
Normalizing the above (equation \ref{sec3:eq:2}) definition by
dividing by the total decay rate we get;
\begin{equation}
{\mathcal A}(s^-, s^+, \hat{s}) = \frac{\overline{\mathcal A}(s^-,
s^+ ,\hat{s})}{d\Gamma/d\hat{s}} . \label{sec3:eq:3}
\end{equation}
We can split the FB asymmetries into their various polarization
components and we will do this in analogy to the prescription given
in Bensalem et al. \cite{Bensalem:2002ni};
\begin{equation}
{\mathcal A}(s^-,s^+) = {\mathcal A} + {\mathcal A}_i s_i^- +
{\mathcal A}_i s_i^+ {\mathcal A}_{ij} s_i^+ s_j^-
\label{sec3:eq:4}
\end{equation}
where $i,j = L,T,N $ are the longitudinal, transverse and normal
components of the polarization\footnote{the convention used here
is that where a repeated index appears it is summed over}.
Therefore, from equation (\ref{sec3:eq:4}) we can write the single
polarized FB asymmetry as;
\begin{eqnarray}
{\mathcal A}_i^- = {\mathcal A}(s^- = i, s^+ = j) + {\mathcal
A}(s^- = i , s^+ = - j) - {\mathcal A}(s^- = - i, s^+ = j) -
{\mathcal A}(s^- = - i , s^+ = - j) \nonumber \\
\label{sec3:eq:5} \\
{\mathcal A}_i^+ = {\mathcal A}(s^- = j, s^+ = i) + {\mathcal
A}(s^- = - j , s^+ = i) - {\mathcal A}(s^- = j, s^+ = - i) -
{\mathcal A}(s^- = - j , s^+ = - i) , \nonumber
\\
\label{sec3:eq:6}
\end{eqnarray}
and the double polarized FB asymmetry as;
\begin{equation}
{\mathcal A}_{ij} = {\mathcal A}(s^- = i, s^+ = j) - {\mathcal
A}(s^- = i , s^+ = -j ) - {\mathcal A}(s^- = -i, s^+ = j ) +
{\mathcal A}(s^- = - i , s^+ = - j) . \label{sec3:eq:7}
\end{equation}
Using the expression of the matrix element as given in equation
(\ref{sec2:eq:2}) we can get the expression of unpolarized FB
asymmetry;
\begin{eqnarray}
{\mathcal A} & = & \frac{3}{\bigtriangleup} \sqrt{1 - \frac{4
\hat{m}_{\ell}^2}{\hat{s}}} \nonumber \\
&&  \times \left[ 2 Re(C_7^{eff} C_{10} ) + \hat{s}
Re(C_9^{eff} C_{10} ) + \hat{m}_{\ell} \left\{Re(C_9^{eff *}
C_{Q_1}) + Re(C_7^{eff *} C_{Q_1}) \right\} \right]
\label{sec3:eq:8}
\end{eqnarray}

\par The analytical results of the polarized FB asymmetries are
then;
\begin{eqnarray}
{\mathcal A}^-_L &=& \frac{1}{\bigtriangleup} \left[
 6 \frac{1}{\hat{s}} |C_7^{eff}|^2 + {3 \over 2} \hat{s} |C_9^{eff}|^2
 + {3 \over 2} \hat{s} \left(1 - \frac{4 \hat{m}_{\ell}^2}{\hat{s}}\right) |C_{10}|^2
 + 6 Re(C_7^{eff *} C_9^{eff}) \right.
 \nonumber \\
&&  \hspace{1cm} - 12 \frac{\hat{m}_{\ell}^2}{\hat{s}} Re(C_7^{eff
*} C_{10}) - 6 \hat{m}_{\ell} Re(C_7^{eff *} C_{Q_2}) - 6
\frac{\hat{m}_{\ell}^2}{\hat{s}} Re(C_9^{eff *} C_{10}) \nonumber \\
&& \hspace{1cm} + 3 \hat{m}_{\ell} Re(C_9^{eff *} C_{Q_2}) \Bigg]
  \label{sec3:eq:9} \\
{\mathcal A}^-_N &=& 0
  \label{sec3:eq:10} \\
{\mathcal A}^-_T &=& \frac{2}{\bigtriangleup} \frac{(1 - \hat{s})
\hat{m}_{\ell}}{\sqrt{\hat{s}}} \sqrt{1 - \frac{4
\hat{m}_{\ell}^2}{\hat{s}}} Re(C_9^{eff *} C_{10})
  \label{sec3:eq:11} \\
{\mathcal A}^+_L &=& - \frac{1}{\bigtriangleup} \left[
 6 \frac{1}{\hat{s}} |C_7^{eff}|^2 + {3 \over 2} \hat{s} |C_9^{eff}|^2
 + {3 \over 2} \hat{s} \left(1 - \frac{4 \hat{m}_{\ell}^2}{\hat{s}}\right) |C_{10}|^2
 + 6 Re(C_7^{eff *} C_9^{eff}) \right.
 \nonumber \\
&&  \hspace{1cm} + 12 \frac{\hat{m}_{\ell}^2}{\hat{s}} Re(C_7^{eff
*} C_{10}) + 6 \hat{m}_{\ell} Re(C_7^{eff *} C_{Q_2}) + 6
\frac{\hat{m}_{\ell}^2}{\hat{s}} Re(C_9^{eff *} C_{10})
    \nonumber \\
&& \hspace{1cm} + 3 \hat{m}_{\ell} Re(C_9^{eff *} C_{Q_2}) \Bigg]
  \label{sec3:eq:12} \\
{\mathcal A}^+_N &=& 0
  \label{sec3:eq:13} \\
{\mathcal A}^+_T &=&
- \frac{2}{\bigtriangleup}
 \frac{(1 - \hat{s}) \hat{m}_{\ell}}{\sqrt{\hat{s}}} \sqrt{1 - \frac{4 \hat{m}_{\ell}^2}{\hat{s}}}
Re(C_9^{eff *} C_{10})
  \label{sec3:eq:14} \\
{\mathcal A}_{LL} &=& \frac{3}{\bigtriangleup} \sqrt{1 - \frac{4
\hat{m}_{\ell}^2}{\hat{s}}} \nonumber \\
&& \hspace{0.7cm} \times \left[ - 2 Re(C_7^{eff} C_{10} ) -
\hat{s} Re(C_9^{eff} C_{10} ) + \hat{m}_{\ell} \left\{Re(C_9^{eff
*} C_{Q_1}) + Re(C_7^{eff *} C_{Q_1}) \right\} \right] \nonumber
\\  \label{sec3:eq:15} \\
{\mathcal A}_{LN} &=& - \frac{2}{\bigtriangleup} \frac{ (1 -
\hat{s}) \hat{m}_{\ell}}{\sqrt{\hat{s}}} \sqrt{1 - \frac{4
\hat{m}_{\ell}^2}{\hat{s}}} Im(C_{10}^* C_9^{eff})
  \label{sec3:eq:16}  \\
{\mathcal A}_{LT} &=&
 \frac{2}{\bigtriangleup } \frac{\hat{m}_{\ell}}{\sqrt{\hat{s}}}
  \left( - 12 \frac{|C_7^{eff}|^2}{\hat{s}} + |C_9^{eff}|^2 \right)
  \label{sec3:eq:17} \\
{\mathcal A}_{NL} &=&
 - {\mathcal A}_{LN}
  \label{sec3:eq:18} \\
{\mathcal A}_{NN} &=&
  - \frac{3}{\bigtriangleup} \hat{m}_{\ell}
  \left( 2 Re(C_7^{eff *} C_{Q_1}) + Re(C_9^{eff *} C_{Q_1}) \right)
 \sqrt{1 - \frac{4 \hat{m}_{\ell}^2}{\hat{s}}}
  \label{sec3:eq:19} \\
{\mathcal A}_{NT} &=&
 \frac{6}{\bigtriangleup} \hat{m}_{\ell}
 \left( 2 \frac{\hat{m}_{\ell}}{\hat{s}} Im(C_{10}^* C_7^{eff})  -
Im(C_7^{eff *}
   C_{Q_2})  + \frac{\hat{m}_{\ell}}{\hat{s}} Im(C_{10}^* C_9^{eff}) \right)
  \label{sec3:eq:20} \\
{\mathcal A}_{TL} &=& - {\mathcal A}_{LT}
  \label{sec3:eq:21} \\
{\mathcal A}_{TN} &=& {\mathcal A}_{NT}
  \label{sec3:eq:22} \\
{\mathcal A}_{TT} &=& - {\mathcal A}_{NN}
  \label{sec3:eq:23}
\end{eqnarray}
where $\bigtriangleup$ is given in equation (\ref{sec2:eq:9}). We
will discuss the above obtained expressions of the various FB
asymmetries and present our numerical analysis of the same in the
next section.


\section{\label{section:4} Numerical analysis and Conclusions}

As it is experimentally more useful we shall present our results
in the form of average values\footnote{it may be possible that the
averaged values of the observables is small but still there could
be large predicted values of the same observables in certain
dileptonic invariant mass regions}. Though, as is well known
\cite{Ali:1996bm}, we have modelled the inclusive decay $B \to X_s
\ell^+ \ell^-$ from a quark level transition $b \to s \ell^+
\ell^-$ using the Heavy Quark Expansion (HQE) in $(1/m_b)$, where
in this expansion we have used the first term of this expansion.
However, this expansion breaks down when $\hat{s} \to 1$. In fact
for higher values of $\hat{s}$ one should consider the
non-perturbative $(1/m_b^2)$ corrections as well
\cite{Ali:1996bm}. As was also shown in \cite{Ali:1996bm} it is
difficult to estimate these non-perturbative corrections for
$\hat{s} \gtrsim  0.8$ . For our analysis we have placed a cutoff
on the upper limit of $\hat{s}$. The averaging procedure which we
will be using is defined as:
\begin{equation}
\langle {\mathcal A} \rangle \equiv
\frac{\displaystyle{\int_{(3.646 + 0.02)^2/m_b^2}^{0.8}} {\mathcal A}
\frac{d \Gamma}{d \hat{s}} d
\hat{s}}{ \displaystyle{\int_{(3.646 + 0.02)^2/m_b^2}^{0.8}} \frac{d
\Gamma}{d \hat{s}} d \hat{s}} ,
\label{sec4:eq:1}
\end{equation}
that is, in the calculation of our averages we have taken the
lower limit of integration to be above the first
resonance\footnote{by first resonance we mean the resonance after
the threshold of the decay, which is $s \geq 4 m_{\tau}^2$.}. Note
that we have used the input parameters presented in appendix
\ref{appendix:1}. The results of table \ref{sec4:tab:1}, these
being our SM predictions of the integrated observables, will of
course be altered when considering a MSSM.

\FIGURE[t]{
\epsfig{file=dr.eps,width=.7\textwidth}
\caption{The variation of
the branching ratio of $B \to X_s \tau^+ \tau^-$ with $tan\beta$
in the mSUGRA model. Other model parameters are: $m_{1/2} =
450$GeV, $A = 200$GeV and the sign of $\mu$ will be taken to be
positive for all our numerical analysis}
\label{dr}
}

\TABULAR{| c | c | c | c | c | c | c | c | c |}
{\hline
Br($B \to X_s \tau^+ \tau^-$)  & ${\mathcal A}$ &
${\mathcal A}^-_L$ & ${\mathcal A}^-_T$
& ${\mathcal A}_{LL}$ & ${\mathcal A}_{LN}$ &
 ${\mathcal A}_{LT}$ &
${\mathcal A}_{NN}$ &
${\mathcal A}_{NT}$  \\
\hline
$1.8 \times 10^{-7}$ & - 0.176 &
0.455 & -0.029 & 0.176 & -0.0084 &
0.063 & 0 & -0.083 \\
\hline}
{Our SM predictions of the integrated observables
\label{sec4:tab:1}}

We have also performed the numerical analysis of the SUSY effects
on various observables, presented in the previous section, for the
inclusive mode ($B \to X_s \tau^+ \tau^-$). If we assume that SUSY
exists, then the MSSM is the simplest SUSY extension of the SM.
However, the MSSM has a large number of parameters, which limits
its practical usage for phenomenological studies. There are many
models which reduce the vast MSSM parameter space to a manageable
set of parameters. These include Dilaton, Moduli, mSUGRA (minimal
Supergravity), rSUGRA (relaxed SUGRA), CMSSM (constraint MSSM)
etc. All of these models have their own characteristics. The basic
feature of all these unified models is that they assume some
unification of the parameters at a higher scale (usually at the
GUT scale). In our numerical analysis we have used one of the more
popular models, minimal Supergravity or mSUGRA\footnote{details of
the model can be found in Nilles \cite{Nilles:1983ge}}.

\FIGURE[ht]{
\epsfig{file=nn_s.eps,width=.7\textwidth}
\caption{$A_{NN}$
variation with the invariant mass of the dileptons, $\hat{s}$, in
the mSUGRA model. The other model parameters are: $m_0 = 400$GeV,
$m_{1/2} = 500$GeV, $A = 0$.}
\label{nn_s}
}

\par In the mSUGRA model one assumes the universality of masses and
coupling constants at the GUT scale. The mSUGRA framework has five
independent parameters\footnote{in Moduli and Dilaton scenarios one
more additional condition is used to reduce the number of
parameters to four} namely, $m_0$ (the unified mass of all the
scalars), $m_{1/2}$ (the unified mass of all the gauginos), $A$
(the unified trilinear coupling constants), $tan\beta$ (the ratio
of vacuum expectation values of the two Higgs doublets). Using
these unified parameters the renormalization group (RG) equations
of all the parameters are evolved from the GUT scale to the
electroweak scale and then at the electroweak scale one imposes
the condition of electroweak symmetry breaking (EWSB). Imposition
of EWSB fixes the magnitude of the parameter $\mu$ \footnote{this
is the coupling parameter of the two Higgs doublets}, but the sign
of $\mu$ still remains arbitrary. As such the $sgn(\mu)$ \footnote{the
convention for $sgn(\mu)$ which we will be going to take is such that
$\mu$ appears in chargino mass matrix with positive sign}
 is taken
to be another parameter, which makes for five parameters (out of
these five, four are continuous parameters with the of $sgn(\mu)$
being discrete, that is $\pm 1$).

\FIGURE[ht]{
\epsfig{file=pol.eps,width=0.9 \textwidth}
\caption{Plots of various
integrated FB asymmetries with the branching ratio of the
inclusive mode ($B \to X_s \tau^+ \tau^-$) in the mSUGRA model.
The mSUGRA parameters are: $m_0 = 500$GeV, $m_{1/2} = 450$GeV, $A
= 0$. In these plots we have varied $tan\beta$ in the range $25
\le tan\beta \le 50$.}
\label{int1}
}

\par Furthermore, we have worked in the high $tan\beta$ region of the
mSUGRA parameter space.  This was done as it is only in this
region that the contributions of the Neutral Higgs Bosons (NHBs)
become significant
\cite{Choudhury:1999ze,Xiong:2001up,RaiChoudhury:1999qb,Choudhury:2003mi,Skiba:1993mg}.
The parameter space of mSUGRA has been constrained by the
experimental observation of $B \to X_s \gamma$ \cite{expbsg}. For
our numerical analysis we will use a more generous bound
\cite{expbsg}:
$$
2 \times 10^{-4} < Br(B \to X_s \gamma) < 4.5 \times 10^{-4}
$$
which is in agreement with CLEO \cite{cleo}, ALEPH \cite{aleph},
BELLE \cite{Abe:2001hk} and BaBar \cite{Aubert:2002pb} results. We
have checked that the qualitative nature of our results doesn't
change even if we use a much stronger bound on the branching ratio
of $B \to X_s \gamma$. As has already been mentioned in many
earlier works \cite{Choudhury:1999ze} it is the purely dileptonic
decay, $B_s \to \mu^+ \mu^-$ which is most affected by the
introduction of the new set of operators. Therefore this decay
would be one of the most promising modes to search for signatures
of these new operators. The SM prediction of this mode is $\sim
10^{-9}$, which would be enhanced by many orders by these new
operators. This process ($B_s \to \mu^+ \mu^-$) has not yet been
observed in B-factories but CDF has given an upper limit on this
branching ratio \cite{nakao}. For our numerical analysis we have
considered only that region of the mSUGRA parameter space which
complies with the new CDF Run II bound $ Br(B_s \to \mu^+ \mu^-) <
9.5 \times 10^{-7} $ \cite{nakao} \footnote{we wish to thank the
referee for updating us on the new CDF result, in our first
version of the manuscript we considered the older CDF bound
\cite{Abe:1998ah} $Br(B_s \to \mu^+ \mu^-) < 2.0 \times 10^{-6}$}.

\par In Figure \ref{dr} we have shown the plots of the branching
ratio of the process concerned as a function of $tan\beta$ for
various values of $m_0$. In Figure \ref{nn_s} we have shown the
variation of the ${\mathcal A}_{NN}$ with the invariant mass of
the dileptons ($\hat{s}$) for various values of $tan\beta$. As can
be seen from the expression of ${\mathcal A}_{NN}$ given in
equation (\ref{sec3:eq:19}) (and also given in Table
\ref{sec4:tab:1}) the asymmetry vanishes within the SM, as within
the SM the scalar and pseudo-scalar operators are negligibly small
($\sim m_\ell m_b/m_W^2$).
Therefore any observation of this asymmetry
can be considered as a signal of new physics. In Figure \ref{int1}
we have shown the values of various integrated FB asymmetries as a
function of the branching ratio. In Figure \ref{cont} we have
shown the contour plots of ${\mathcal A}_{NN}$ in the $[m_0,m_{1/2}]$
plane for two different values of $tan\beta$.

\par As has been mentioned in many earlier works
\cite{Choudhury:1999ze,Xiong:2001up,RaiChoudhury:1999qb,Choudhury:2003mi,Skiba:1993mg},
the effect of the scalar (and pseudo-scalar) operators crucially
depends on $tan\beta$ and the mass of the Higgs. One could have
used some relaxed kind of SUGRA framework also. By relaxed we mean
relaxing the universality of the masses and coupling constants.
Here we can relax the conditions of universality of scalar and/or
gaugino masses. In the literature these models have been termed as
rSUGRA (or relaxed SUGRA models) \cite{goto1}. In these situations
one can have sufficiently low allowed\footnote{by allowed we mean that
values are consistent with experimental constraints.} values
of the Higgs mass for large $tan\beta$. In such situations the
numerical values of the Wilson coefficients accompanying the
scalar and pseudo-scalar operators could be high, and hence their
effects on observables could be more profound. Similarly one could
have used CMSSM models, where even the condition of correct
electroweak symmetry breaking is relaxed. In these models one
can have a very large variation in the Wilson coefficients
$C_{Q_1}$ and $C_{Q_2}$. However, in this work we have confined
ourselves to the mSUGRA model.

\FIGURE[h]{
\epsfig{file=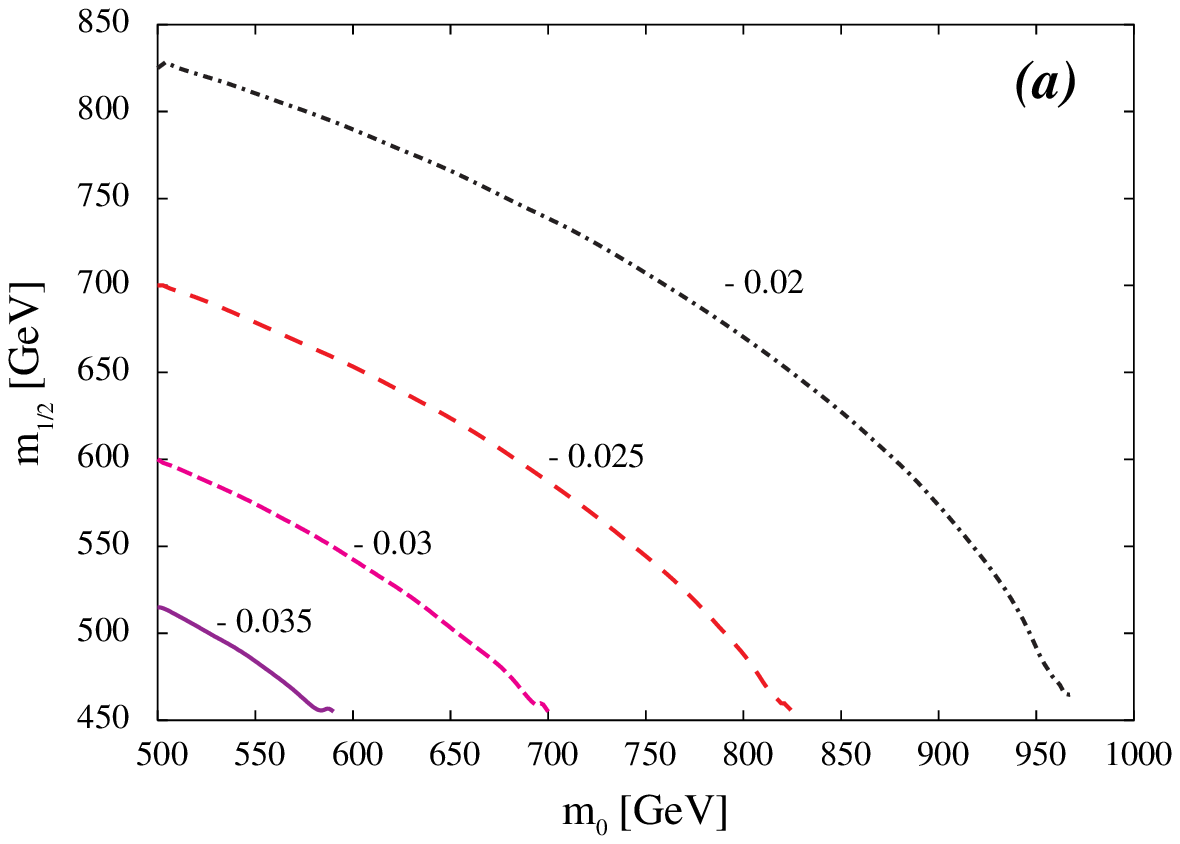,width=.48\textwidth}
\epsfig{file=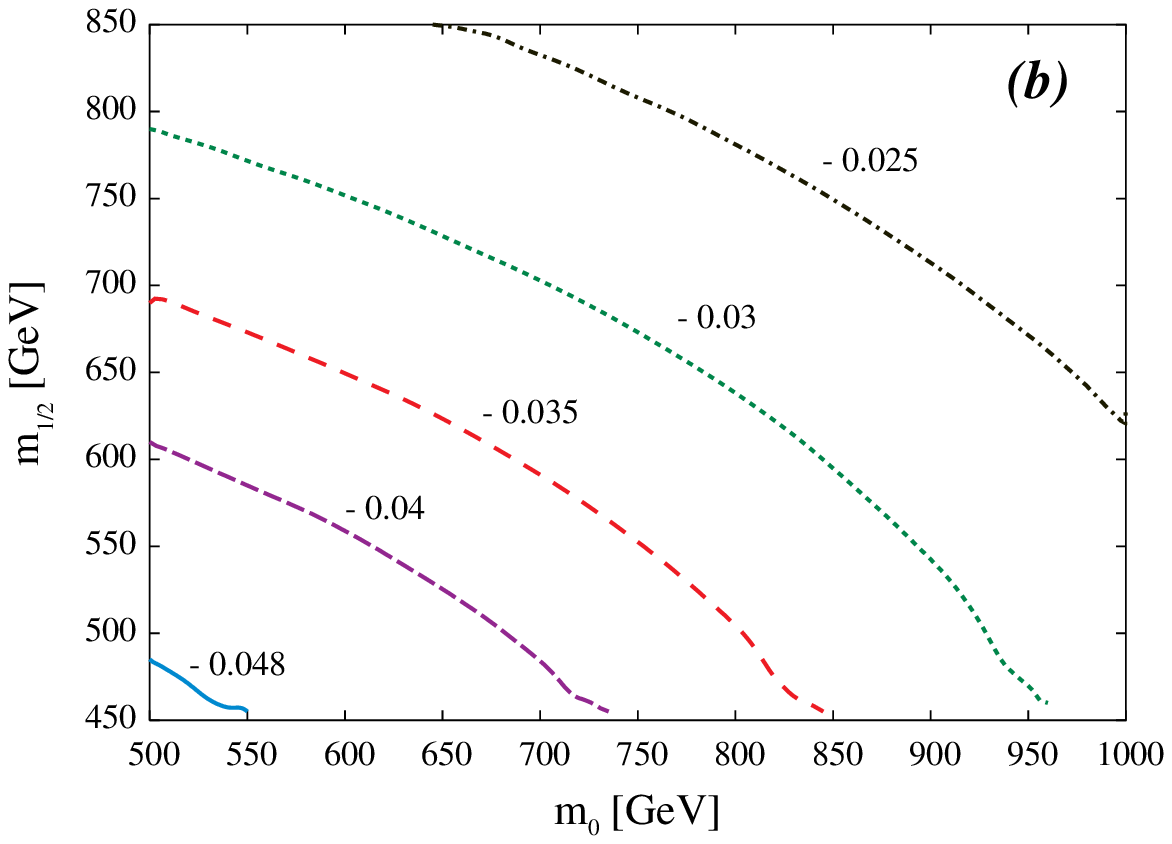,width=.48\textwidth}
\caption{Contour plots of
various values of averaged $<{\mathcal A}_{NN}>$ (as given in plots)
in the $[m_0,m_{1/2}]$ plane for the mSUGRA model with $A = 0$ and
{\bf (a)} $tan\beta = 40$, {\bf (b)} $tan\beta = 45$.}
\label{cont}
}

\par So far no experimental search for the inclusive decay mode $B \to
X_s \tau^+ \tau^-$ has been carried out. Experimentally this decay
can be searched for by observing the $\mu$ produced from the
$\tau$ via the $\tau \to \mu \nu \bar{\nu}$ process, but the muons
produced in this cascade decay would be soft ones (so to be
observed one has to observe muon pairs at low invariant mass). But
there was an upper bound put on the branching ratio of this mode
($B \to X_s \tau^+ \tau^-$) by Grossman et al.
\cite{Grossman:1996qj} which was:
$$
Br(B \to X_s \tau^+ \tau^-) < 5 \% .
$$
As the $\tau$ detection efficiency is low, such that at the SM
level this decay mode might remain out of reach of the present
generation B-factories (like BaBar, BELLE etc.), it has been well
emphasized in many earlier works
\cite{Guetta:1997fw,Grossman:1996qj,Xiong:2001up} that this decay
can be enhanced by a couple of orders in magnitude in the presence
of new physics and hence could be very useful in putting bounds on
new physics models.

\par To summarize the major conclusions of our numerical analysis:
\begin{enumerate}
\item[1.] Almost all the polarized asymmetries show a large variation
from their respective SM values over a large parameter space at
large $tan\beta$.
\item[2.] The SM relationship that the magnitude of ${\mathcal A}_{LL}$
is the same as ${\mathcal A}$ is violated in the presence of new
operators.
\item[3.] The observation of a non-vanishing value for ${\mathcal
A}_{NN}$ could be treated clearly as a signal of new physics.
\end{enumerate}

These polarized FB asymmetries also provide us with much needed
observables, which could aid in the testing of the effective
Hamiltonian's structure; especially the scalar and pseudo-scalar
parts as can be see from the expression for ${\mathcal A}_{NN}$
which is proportional to $C_{Q_1}$ and hence vanishes within the SM.


\acknowledgments The authors wish to thank Prof. S. Rai Choudhury
for his useful discussions during the course of this work. The
authors would also like to thank Z. Ligeti and Y. Grossman for
their useful communication.  This work of N.G. was supported under
the SERC scheme of the Department of Science \& Technology (DST),
India.


\appendix

\section{\label{appendix:1} Input parameters}

\begin{center}
$m_b ~=~ 4.8$ GeV  \ , \
$ m_c ~=~ 1.4 $ GeV  \\
$m_\tau ~=~ 1.77$  GeV  \ , \
$m_w ~=~ 80.4$  GeV  \ , \\
$m_z ~=~ 91.19$  GeV  \ , \
$V_{tb} V^*_{ts} = 0.0385$  , \\
$\alpha = {1 \over 129}$   \ , \ $G_F = 1.17 \times 10^{-5} ~{\rm
GeV}^{-2}$
\end{center}


\end{document}